\documentclass[prl,twocolumn,showpacs,amsmath,amssymb]{revtex4-1}
\usepackage{subfigure}
\usepackage{graphicx}
\usepackage{dcolumn}
\usepackage{float}
\usepackage{xcolor}
\usepackage{epstopdf}
\usepackage{makecell}
\usepackage{overpic}
\usepackage{multirow}
\usepackage{enumerate}
\usepackage{lineno}
\usepackage{hyperref}

\newcommand*{\LamC}{\ensuremath{\Lambda_{c}^+}}
\newcommand*{\ALamC}{\ensuremath{\bar{\Lambda}_{c}^{-}}}
\newcommand*{\LamCF}{\ensuremath{\Lambda_{c}(2595)^+}}
\newcommand*{\LamCS}{\ensuremath{\Lambda_{c}(2625)^+}}
\newcommand*{\ALamCF}{\ensuremath{\bar{\Lambda}_{c}(2595)^-}}
\newcommand*{\ALamCS}{\ensuremath{\bar{\Lambda}_{c}(2625)^-}}
\newcommand*{\LCLC}{\ensuremath{\Lambda_{c}^+\bar{\Lambda}_{c}^{-}}}
\newcommand*{\Mpkpi}{\ensuremath{M_{p K^{-}\pi^{+}}}}
\newcommand*{\Mrec}{\ensuremath{M_{\LamC}^{\rm rec}}}
\newcommand*{\eeLLF}{\ensuremath{e^+e^-\to \LamC \bar{\Lambda}_{c}(2595)^-}}
\newcommand*{\eeLLS}{\ensuremath{e^+e^-\to \LamC \bar{\Lambda}_{c}(2625)^-}}
\newcommand*{\AeeLLF}{\ensuremath{e^+e^-\to \ALamC \Lambda_{c}(2595)^+}}
\newcommand*{\AeeLLS}{\ensuremath{e^+e^-\to \ALamC \Lambda_{c}(2625)^+}}
\newcommand{\ee}{\ensuremath{e^+e^-}}
\newcommand{\ffratio}{\sqrt{|G_{E}|^2 + 3|G_{M}|^2}/|G_{C}|}
\newcommand{\gev}{\mathrm{GeV}}
\newcommand{\mev}{\mathrm{MeV}}
\newcommand{\gevcc}{\mathrm{GeV}/c^2}

\begin{document}
\title{Measurements of Born Cross Sections for $\eeLLF+\rm{c.c.}$ and $\eeLLS+\rm{c.c.}$ at $\sqrt{s}=$~4918.0 and 4950.9~MeV}

%%\title{Measurements of cross section of $\eta\phi$ production above 3.773 GeV}
%\title{Cross section Measurements of $\EE \to \ep$ at $\sqrt s=$ 3.773 to 4.600 GeV }
\author{M.~Ablikim$^{1}$, M.~N.~Achasov$^{4,b}$, P.~Adlarson$^{75}$, O.~Afedulidis$^{3}$, X.~C.~Ai$^{80}$, R.~Aliberti$^{35}$, A.~Amoroso$^{74A,74C}$, Q.~An$^{71,58}$, Y.~Bai$^{57}$, O.~Bakina$^{36}$
% ...
\\
\vspace{0.2cm}
(BESIII Collaboration)\\
\vspace{0.2cm} {\it
$^{1}$ Institute of High Energy Physics, Beijing 100049, People's Republic of China\\
% ...
$^{3}$ Bochum  Ruhr-University, D-44780 Bochum, Germany\\
$^{4}$ Budker Institute of Nuclear Physics SB RAS (BINP), Novosibirsk 630090, Russia\\
% ...
$^{35}$ Johannes Gutenberg University of Mainz, Johann-Joachim-Becher-Weg 45, D-55099 Mainz, Germany\\
$^{36}$ Joint Institute for Nuclear Research, 141980 Dubna, Moscow region, Russia\\
% ...
$^{57}$ Southeast University, Nanjing 211100, People's Republic of China\\
$^{58}$ State Key Laboratory of Particle Detection and Electronics, Beijing 100049, Hefei 230026, People's Republic of China\\
% ...
$^{71}$ University of Science and Technology of China, Hefei 230026, People's Republic of China\\
% ...
$^{74}$ University of Turin and INFN, (A)University of Turin, I-10125, Turin, Italy; (B)University of Eastern Piedmont, I-15121, Alessandria, Italy; (C)INFN, I-10125, Turin, Italy\\
$^{75}$ Uppsala University, Box 516, SE-75120 Uppsala, Sweden\\
% ...
$^{80}$ Zhengzhou University, Zhengzhou 450001, People's Republic of China\\
\vspace{0.2cm}
% ...
$^{b}$ Also at the Novosibirsk State University, Novosibirsk, 630090, Russia\\
% ...
}}

\vspace{0.4cm}

\begin{abstract}
Using $e^+e^-$ collision data collected with the BESIII detector operating at the BEPCII collider, the Born cross sections of $\eeLLF+\rm{c.c.}$ and $\eeLLS+\rm{c.c.}$ are measured for the first time at center-of-mass energies of $\sqrt{s}=4918.0$ and $4950.9~\mev$.
Non-zero cross sections are observed very close to the production threshold.
The measured Born cross sections of $\eeLLS+\rm{c.c.}$ are about $2\sim3$ times greater than those of $\eeLLF+\rm{c.c.}$, providing the similar behavior as semileptonic decays of $\Lambda_{b}^0$, but different behavior from that in the hadronic decays of $\Lambda_{b}^0$. The Born cross sections are $15.6\pm3.1\pm0.9$ pb and $29.4\pm3.7\pm2.7$ pb for $\eeLLF+\rm{c.c.}$, and are $43.4\pm4.0\pm4.1$ pb and $76.8\pm6.5\pm4.2$ pb for $\eeLLS+\rm{c.c.}$ at $\sqrt s=4918.0$ and 4950.9 MeV, respectively.
Based on the polar angle distributions of the $\ALamCS$ and $\LamCS$, the form-factor ratios $\ffratio$ are determined for $\eeLLS+\rm{c.c.}$ for the first time, which are $5.95\pm4.07\pm0.15$ and $0.94\pm0.32\pm0.02$ at $\sqrt s=4918.0$ and 4950.9 MeV, respectively. All of these first uncertainties are statistical and second systematic.
\end{abstract}

%\linenumbers
\maketitle

Since the discovery of the ground-state $\LamC$~\cite{Belle:2013jfq1,Belle:2013jfq}, abundant excited charmed baryons have been reported in experiment~\cite{PDG:2022}.
Studies of their spectroscopy and internal structures provide important information for the understanding of the chiral symmetry breaking and the heavy quark symmetry~\cite{Cheng:2021qpd}.
The first two excited states of $\LamC$ are $\LamCF$ and $\LamCS$~\cite{Argus:LamCstar1,Argus:LamCstar2,CLEO:LamCstar,E687:LamCstar}.
Under the prediction of the quark model, they correspond to the degenerate pair of $P$-wave excited states, with spin-parity 1/2$^-$ and 3/2$^-$, respectively.
In recent years, their exotic properties have attracted much attention in both theory and experiment.

The CDF collaboration reported the production of $\LamCF$ and $\LamCS$ in the semileptonic decays of $\Lambda^{0}_{b}$~\cite{CDF:2008hqh}.
The measured decay rate of $\Lambda_{b}^{0}\to\LamCS\mu^-\bar{\nu}_{\mu}$ is about twice that of $\LamCF\mu^-\bar{\nu}_{\mu}$, which contradicts the calculation of lattice QCD (LQCD) in the larger $q^2$ region~\cite{Meinel:2021rbm,Meinel:2021mdj}.
Based on the conventional quark configuration, LQCD calculation gives a much lower differential decay rate of $\Lambda_{b}^{0}\to\LamCS\mu^-\bar{\nu}_{\mu}$.
Because the measured mass of $\LamCF$ lies just on the threshold of $\Sigma_{c}(2455) \pi$ within a few MeV~\cite{PDG:2022},
the exotic structure of $\LamCF$ was proposed to interpret the contradiction, e.g., the dynamically generated meson-baryon states~\cite{Lu:exotic,Nieves:exotic} that is analogous to the reverse problem of $\Lambda(1405)$~\cite{CLAS:2014tbc} and $\Lambda(1520)$~\cite{Kamano:2015hxa}.
The LHCb collaboration reported the production of $\LamCF$ and $\LamCS$ in the hadronic decays of $\Lambda^{0}_{b}$~\cite{LHCb:2011poy}.
The measured decay rates of $\Lambda_{b}^{0}\to\LamCS\pi^-$ and $\Lambda_{b}^{0}\to\LamCF\pi^-$ are equivalent, which provides different behavior from that in the semileptonic decays.
So far, the nature of $\LamCF$ and $\LamCS$ is still mysterious and experimental results are limited.
The study of baryon-pair production via $\ee$ annihilation is an effective and crucial approach to disentangle the structure information which is always involved in the theoretical calculation for the interaction vertex between virtual photon and baryons.
At the moment, there are no such experimental results for these two excited charmed baryons.
Therefore, the measurements of the production cross sections for the processes involving these excited charmed baryons are essential to disentangle different interpretations.

In the center-of-mass energy range from 2.0  to 4.95~$\gev$, BESIII has already performed a series of measurements, not only production cross sections of baryon pairs, but also their electromagnetic form factors (EMFFs)~\cite{Huang:2021xte,BESIII:protonFF,BESIII:neutronFF,BESIII:2023ioy,BESIII:SigmaC,BESIII:Sigma0,BESIII:LcLc,BESIII:XiXi,BESIII:OmegaOmega,BESIII:LcLcnew}. Above $4.9~\gev$, the production of $\LamCF$ or $\LamCS$ accompanied by the ground-state $\LamC$ is allowed.
The data samples above 4.9 GeV~\cite{BESIII:Lumi} provide an ideal opportunity to study the production of $\ee\to\LamC\ALamCF+\rm{c.c.}$ and $\ee\to\LamC\ALamCS+\rm{c.c.}$.

In this Letter, we report the measurements of the Born cross sections of the processes $\eeLLF+\rm{c.c.}$ and $\eeLLS+\rm{c.c.}$ processes
at the two center-of-mass (c.m.)~energies of $\sqrt{s}= 4918.0$ and $4950.9~\mev$ for the first time.
The integrated luminosities at these energy points are $207.8\pm1.1$ and $159.3\pm0.9$~pb$^{-1}$~\cite{BESIII:Lumi}, respectively.
Additionally, we analyze the polar angle distributions of $\ALamCS$ and its antiparticle, $\LamCS$, to determine the form-factor ratios $\ffratio$ for the first time near the production threshold of $\eeLLS+\rm{c.c.}$.

A detailed description of the design and performance of the BESIII detector can be found in Ref.~\cite{Ablikim:2009aa,Ablikim:2019hff}.
Simulated events, which are produced with the {\sc geant4}-based~\cite{Agostinelli:2002hh} Monte Carlo (MC) simulation programs implementing the geometric description~\cite{detvis} of the BESIII detector and detector response, are used to determine detection efficiencies and to estimate backgrounds.
The signal MC samples of the $\eeLLF$ and $\eeLLS$ processes, where $\LamC$ directly from $\ee$ annihilation decays into $p K^-\pi^+$ and $\ALamCF/\ALamCS$ decays generically, are simulated at individual c.m.~energies using the {\sc kkmc} generator~\cite{Jadach:1999vf}.
The software package {\sc besevtgen}~\cite{Ping:2008zz} handles the procedure of subsequent decays after the productions of $\eeLLF$ and $\eeLLS$.
These samples are denoted as the $S_{ee}$ contribution in this analysis.
Meanwhile, another signal MC samples for the processes $\AeeLLF$ and $\AeeLLS$ are also produced, where the $\LamC$ decaying into $p K^{-}\pi^+$ comes from the excited charmed baryon decays, i.e, $\LamCF$ and $\LamCS$,  while $\ALamC$ decays generically.
These samples are denoted as $S_{\rm inte}$.
The charge-conjugate processes are also generated and included in $S_{ee}$ and $S_{\rm inte}$.
%According to the Particle Data Group (PDG)~\cite{PDG:2022}, the $\Lambda_{c}(2595)^{+}$ and $\Lambda_{c}(2625)^{+}$ eventually decay to the final states of $\Lambda_{c}^{+}\pi^+\pi^-$ and $\Lambda_{c}^{+}\pi^0\pi^0$, where the intermediate decay $\Lambda_{c}(2595)^{+}\to\Sigma_{c}(2455)\pi$ takes a fraction of 67\%, and the ratio between $\LamC\pi^+\pi^-$ and $\LamC\pi^0\pi^0$ is 2:1.
The $\Lambda_{c}(2595)^{+}$ and $\Lambda_{c}(2625)^{+}$ eventually decay to the final states of $\Lambda_{c}^{+}\pi^+\pi^-$ and $\Lambda_{c}^{+}\pi^0\pi^0$, and the ratio between $\LamC\pi^+\pi^-$ and $\LamC\pi^0\pi^0$ is 2:1 as the expectation under isospin assumption~\cite{ARGUS:1993vtm}.
But as reported in the Particle Data Group (PDG)~\cite{PDG:2022}, the indirect determination this ratio~\cite{CDF:2011zbc} can violate the naive expectation.
For all these signal MC samples, the initial-state radiation (ISR)~\cite{Jadach:2000ir} and the beam energy spread~\cite{Brodsky:2009gx} are implemented during the generation process.
In addition, the c.m. energy-dependent Born cross sections, measured and parameterized in this work, are inputs in the {\sc kkmc} generator iteratively.
To achieve a better simulation, the polar angle ($\theta$) distributions of $\ALamCF$ and $\ALamCS$ are considered in the generator via a parametrization of $f(\cos\theta) \propto (1+\alpha_{\Lambda_{c}}\!\cos^2\!\theta)$.
For the $\eeLLS$ production, the value of $\alpha_{\Lambda_{c}}$ is assigned as the one measured by this analysis, while for $\eeLLF$,  $\alpha_{\Lambda_{c}}=1$ is used, and possible deviation from it is considered
as a source of systematic uncertainty.
The angular distributions are also taken into account in the generator for the charge-conjugate processes.
To study the background, the inclusive MC samples, including the $\ee\to\LCLC$ events, the $\ee\to\ell^{+}\ell^{-}$ ($\ell=e$, $\mu$, and $\tau$) events, the $D_{(s)}$ production, the ISR return to lower-mass $\psi$ states, and the continuum processes $\ee\to q\bar{q}$ ($q=u,d,s$), are produced.
The final-state radiation of charged final-state particles is simulated using the {\sc photos}~\cite{Richter-Was:1992hxq} package.
All these inclusive background MC samples, except for the $\ee\to\ell^{+}\ell^{-}$ events, are combined according to the corresponding observed cross sections and referred to as the $q\bar{q}$ process hereafter.
Unless explicitly stated, the charge-conjugate processes are implied in the following description of selecting signal candidates.

%\vspace{-1.25em}
\begin{figure*}[!htbp]
\begin{center}
\hspace{-6.0em}
 \subfigure{\includegraphics[width=0.56\textwidth,height=0.35\textwidth]{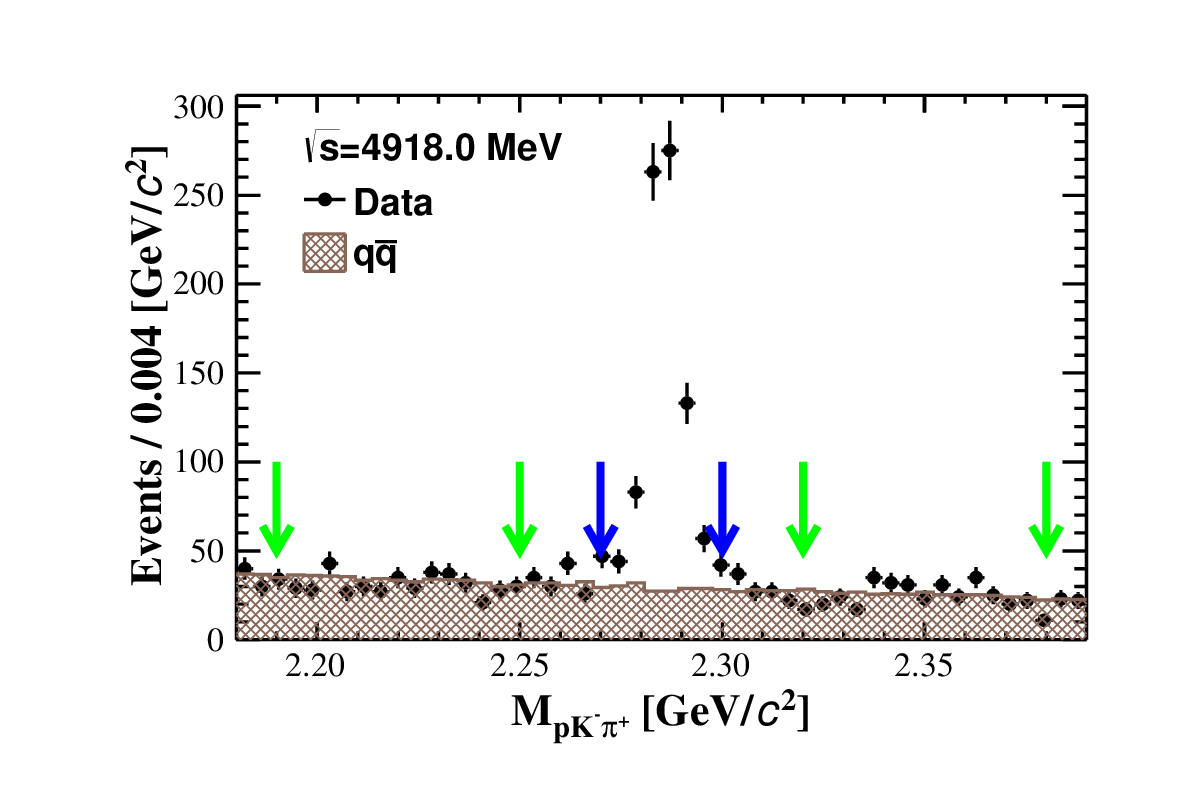}\label{data:Lc4918}}\put(-83,125){ ~\bf (a)}
 \hspace{-1.5em}
\subfigure{\includegraphics[width=0.56\textwidth,height=0.35\textwidth]{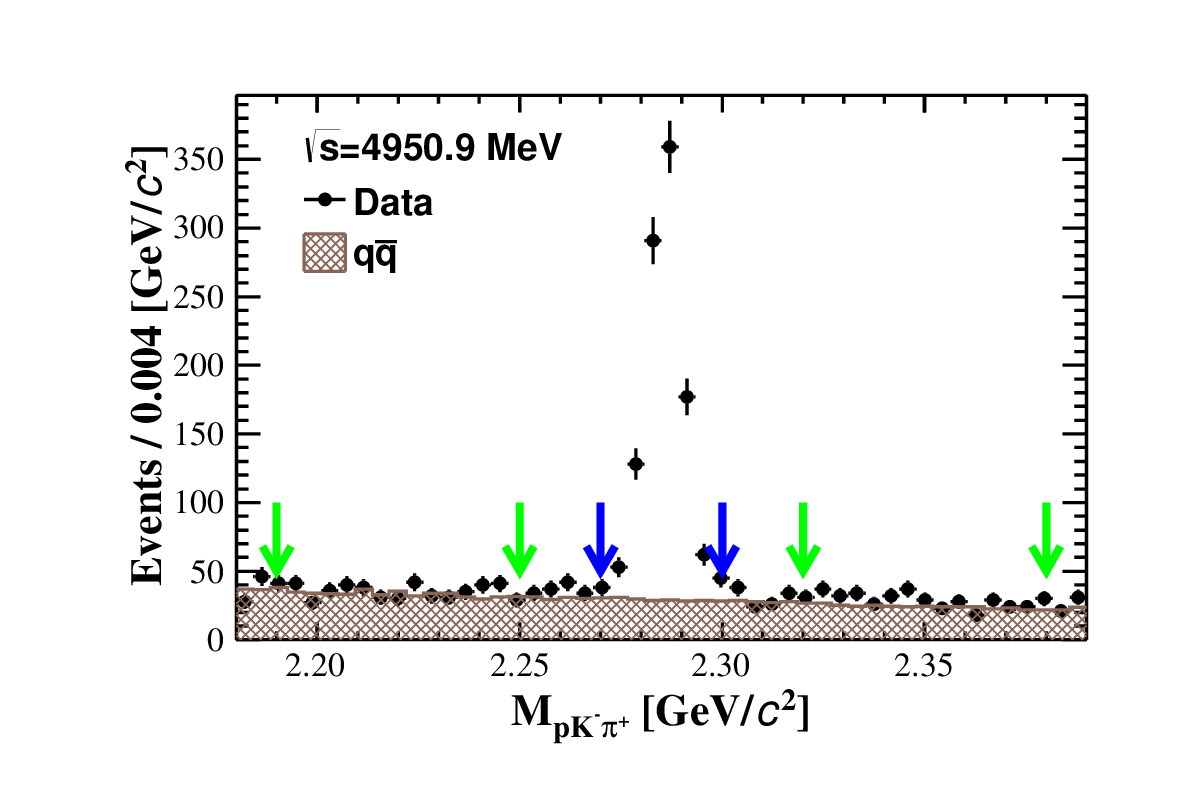}\label{data:Lc4946}}\put(-83,125){ ~\bf (b)}
\hspace{-5.8em}
\setlength{\abovecaptionskip}{-0.4em}
\caption{The $\Mpkpi$ distributions of the $\LamC$ candidates at $\sqrt s=$~(a)~4918.0 and (b) 4950.9 MeV, where the shaded histograms is the $q\bar{q}$ contribution derived from the inclusive MC samples. The region between the blue arrows is the signal region, and the regions between two neighbor green arrows are the sideband regions.}
\label{fig::mrecLam0}
\end{center}
\end{figure*}
%\vspace{-1.2em}

\begin{figure*}[!htb]
\begin{center}
\hspace{-6.0em}
\subfigure{\includegraphics[width=0.55\textwidth,height=0.33\textwidth]{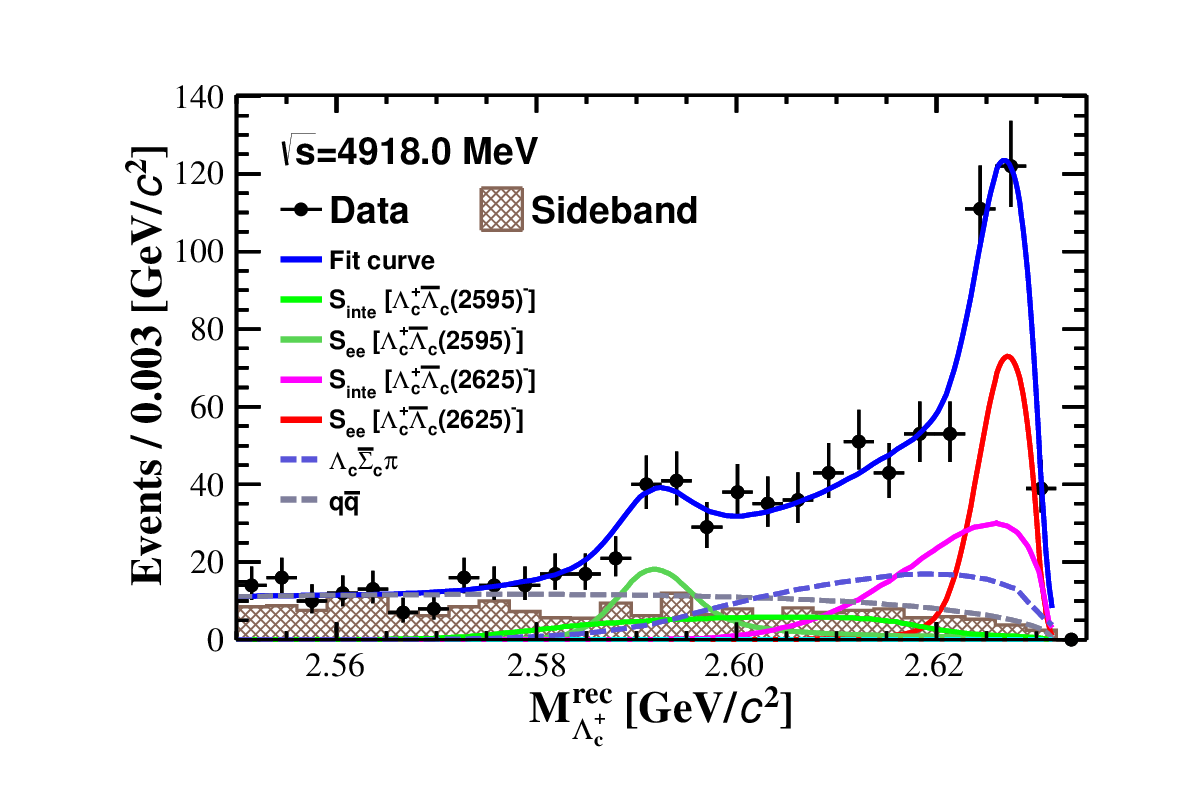}\label{fit:4918}}\put(-83,125){ ~\bf(a)}
\hspace{-1.5em}
\subfigure{\includegraphics[width=0.55\textwidth,height=0.33\textwidth]{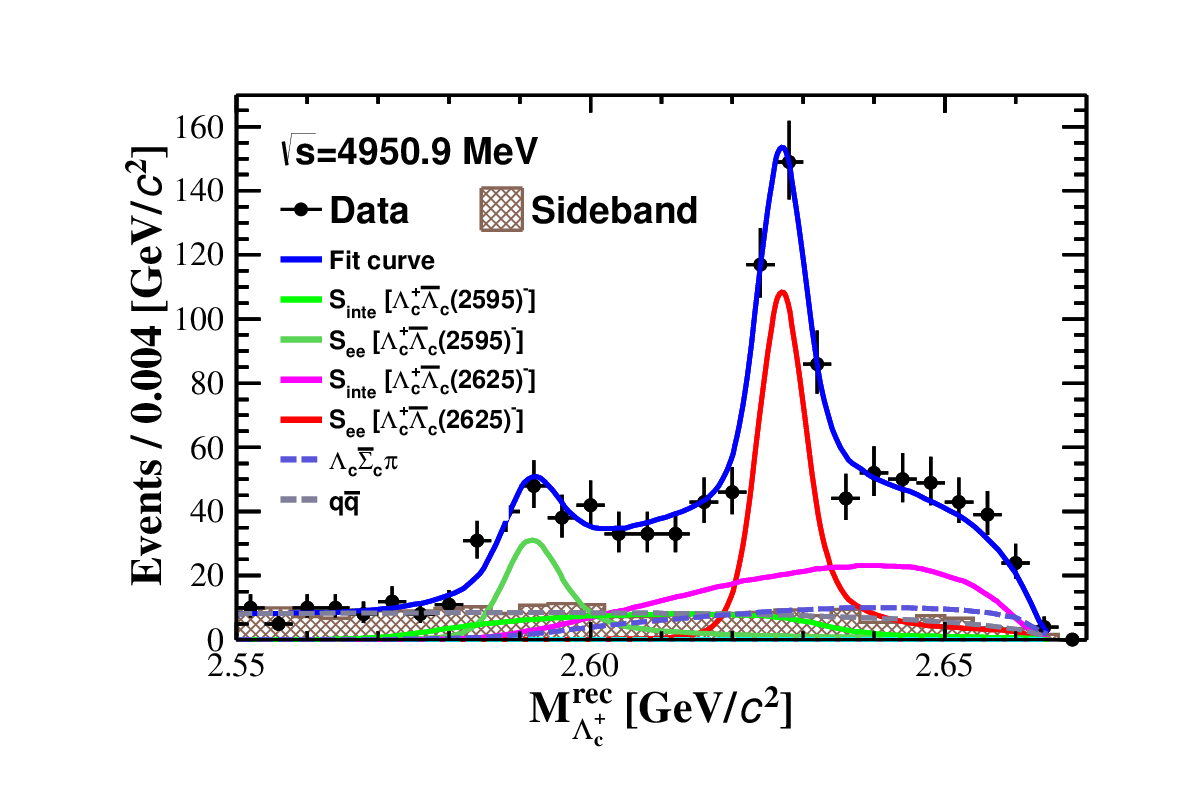}\label{fit:4946}}\put(-83,125){  ~\bf(b)}
\hspace{-5.8em}
\setlength{\abovecaptionskip}{-0.4em}
\caption{The recoiling mass distributions $\Mrec$ of the $\LamC$ candidates at $\sqrt s=$~(a)~4918.0 and (b) 4950.9 MeV, where the shaded histograms are the events in the $\Mpkpi$ sideband regions in data.
The blue solid lines are the total fit curves, including the signal contributions of $\eeLLF+\rm{c.c.}$ and $\eeLLS+\rm{c.c.}$, and the background contributions of $q\bar{q}$, $\ee\to \Sigma_{c}\bar{\Sigma}_{c}$, and $\ee\to\LamC\bar{\Sigma}_{c}\pi+\rm{c.c.}$. The blue dashed lines are the contributions of $\ee\to\LamC\bar{\Sigma}_{c}\pi+\rm{c.c.}$.
Other lines label individual ones.
}\label{fig::mrecLam}
\end{center}
\end{figure*}

The $\LamC$ candidates are selected with the following criteria:
(i) the charged tracks detected in the helium-based main drift chamber (MDC) must satisfy $|\!\cos\theta|<0.93$ where $\theta$ is defined with respect to the $z$-axis, which is the symmetry axis of the MDC, and have a distance of closest approach to the interaction point of less than 10~cm along the $z$-axis and less than 1~cm in the perpendicular plane;
(ii) the particle identification (PID) is implemented by calculating the probability using the specific ionization energy loss information provided by the MDC and the time-of-flight information, and each charged track is assigned a particle type hypothesis ($p$, $K^{-}$, and $\pi^{+}$) with the highest probability.
To avoid losing the signal events, all the $p K^-\pi^+$ combinations are retained.
The distributions of the $pK^-\pi^+$ invariant masses, $\Mpkpi$, after requiring $\Mrec>2.55~\gevcc$, are shown in Figs.~\ref{fig::mrecLam0}(a) and \ref{fig::mrecLam0}(b), where evident $\LamC$ signals are observed at both energy points.
The $\Mrec$ is the recoiling mass against the reconstructed $\LamC$ in the c.m. system.
After requiring $\Mpkpi\in(2.27,~2.30)~\gevcc$, clear $\ALamCF$ and $\ALamCS$ signals are observed in the $\Mrec$ spectra, as illustrated in
Figs.~\ref{fig::mrecLam}(a) and \ref{fig::mrecLam}(b), which indicate the existence of the processes $\eeLLF+\rm{c.c.}$ and $\eeLLS+\rm{c.c.}$.
Studies based on the inclusive MC samples show that the dominant background contamination is from the $q\bar{q}$ process.

%\vspace{-1.2em}
%\vspace{-1.2em}

The Born cross sections ($\sigma$) for individual signal processes are obtained with unbinned maximun-likelihood fits
to the $\Mrec$ distributions. In the fit, the total yield $N_{\rm obs}$ takes
\begin{equation}\label{eq:xsec}
\begin{aligned}
N_{\rm obs} = N_{\rm sig}^{2595} + N_{\rm sig}^{2625} + N_{\rm bkg},
\end{aligned}
\end{equation}
where $N_{\rm sig}^{2595}$ and $N_{\rm sig}^{2625}$ are the yields of the $\eeLLF+\rm{c.c.}$ and $\eeLLS+\rm{c.c.}$ processes, respectively, while $N_{\rm bkg}$ is the number of total residual background events. The signal yields for each process are related with the Born cross sections $\sigma$  by
\begin{equation}\label{eq:yield}
\begin{aligned}
N_{\rm sig}  = \sigma\mathcal{L}f_{\rm VP} f_{\rm ISR}\mathcal{B} (\varepsilon_{ee}+\varepsilon_{\rm inte}),
\end{aligned}
\end{equation}
where $\mathcal{L}$ is the integrated luminosity, $f_{\rm ISR}$ is the ISR correction factor derived from the QED theory~\cite{Jadach:2000ir} and calculated by inputting the line shape of the Born cross sections into the generator, $f_{\rm VP}=1.06$ is the vacuum polarization (VP) correction factor~\cite{HC:1985gq,WGRC:2010bjp}, and $\mathcal{B}=(6.26\pm0.29)\%$ is the branching fraction of $\LamC\to pK^{-}\pi^{+}$~\cite{PDG:2022}.
The $\varepsilon_{ee}$ stands for the detection efficiency of $S_{ee}$ contributions and $\varepsilon_{\rm inte}$ for $S_{\rm inte}$ contributions. There should exist factor 1/2 in Eq.~\ref{eq:yield}, accounting for that the reconstructed $\LamC\to pK^{-}\pi^+$ has been categorized to be either directly from $\ee$ collision or decayed particle of the intermediate states, which is canceled due to the charge-conjugate processes have been taken into account.
The charge-conjugate processes have been assumed to have equal Born cross sections.
The signal shapes are derived from the signal MC samples, and those of the $S_{ee}$ contributions are convolved with Gaussian functions accounting for the differences between data and MC simulation.
The same Gaussian functions are shared between the $\ALamCF$ and $\ALamCS$ signals due to the limited statistics.
The $q\bar{q}$ contributions are described with an ARGUS function~\cite{ARGUS:1990hfq}, whose parameters are determined by fitting the events in the $\Mpkpi$ sideband regions of $\Mpkpi\in(2.19,\, 2.25)$ and $(2.32,\, 2.38)~\gevcc$, as indicated by the green arrows in Figs.~\ref{fig::mrecLam0}(a) and \ref{fig::mrecLam0}(b).
Moreover, at both energy points, the $\ee\to \Sigma_{c}\bar{\Sigma}_{c}$ and $\ee\to\LamC\bar{\Sigma}_{c}\pi+\rm{c.c.}$ (${\Sigma}_{c}={\Sigma}_{c}^0$, ${\Sigma}_{c}^{+}$, and ${\Sigma}_{c}^{++}$) processes potentially exist, as they have the same final state as the signal processes.
Both processes are taken into account as additional background components, whose shapes are modeled by the MC simulation.
Due to the unknown production cross sections of these two background processes, their normalization factors are free in the fit.
The total number of the background events, {\it i.e.}, $N_{\rm bkg}$, includes the contributions from the $q\bar{q}$, $\ee\to\Sigma_{c}\bar{\Sigma}_{c}$, and $\ee\to\LamC\bar{\Sigma}_{c}\pi+\rm{c.c.}$ processes.
The fit results are illustrated in Figs.~\ref{fig::mrecLam}(a) and \ref{fig::mrecLam}(b) and listed in Table~\ref{tab::XS}.
Due to the yields of process  $\ee\to \Sigma_{c}\bar{\Sigma}_{c}$ are negligible at both energy points, according to the fitting results, only the contributions of process $\ee\to\LamC\bar{\Sigma}_{c}\pi+\rm{c.c.}$ are shown in Figs~\ref{fig::mrecLam} .

\begin{figure*}[!htbp]
\begin{center}
\hspace{-6em}
\subfigure{\includegraphics[width=0.55\textwidth,height=0.33\textwidth]{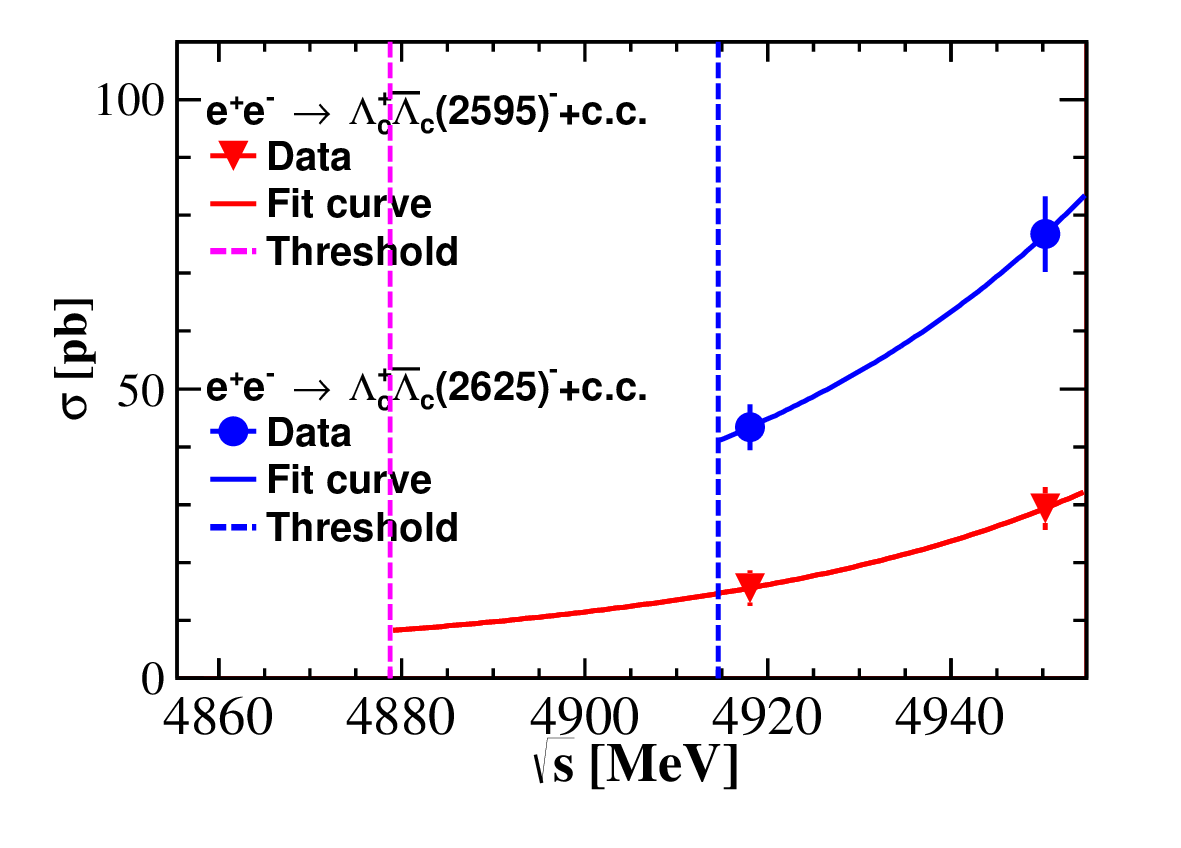}\put(-83,125){ ~\bf(a)}\label{result:xsec}}
\hspace{-1.3em}
\subfigure{\includegraphics[width=0.55\textwidth,height=0.33\textwidth]{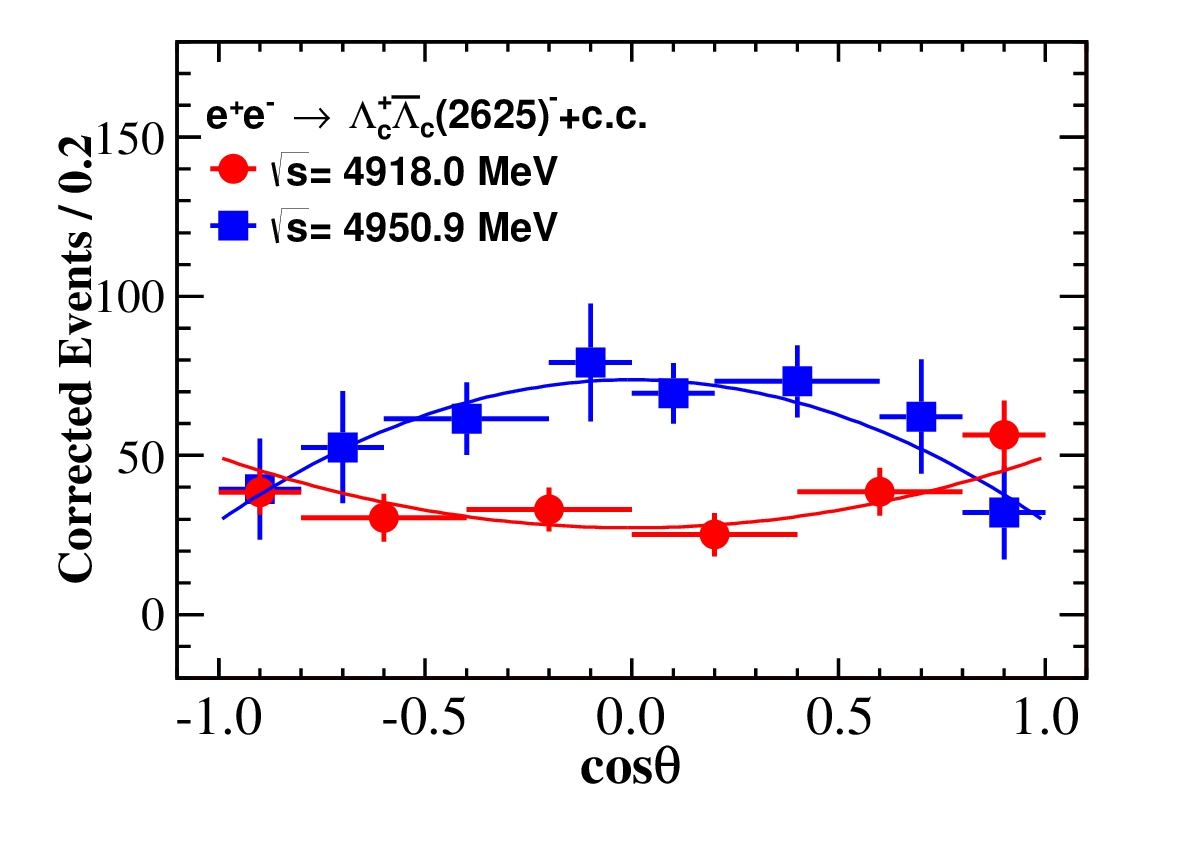}\put(-83,125){ ~\bf(b)}\label{result:alpha}}
\hspace{-5.8em}
\setlength{\abovecaptionskip}{-0.4em}
\caption{(a) The Born cross sections of $\eeLLF+\rm{c.c.}$ and $\eeLLS+\rm{c.c.}$ and the fits using Eq.~(\ref{eq:method1}).
(b) The  efficiency-corrected polar angle distributions of $\eeLLS+\rm{c.c.}$ and the fits with the $f(\cos\theta)$ function.}
\label{fig::Line_alpha}
\end{center}
\end{figure*}

Since $f_{\rm ISR}$ and detection efficiencies depend on the input Born cross sections line shape, an iterative approach is employed to obtain stable $f_{\rm ISR}$ and detection efficiencies. A perturbative QCD-motivated energy power function~\cite{Pacetti:2014jai,Lepage:1980fj} is used to model the Born cross sections line shape as
\begin{equation} \label{eq:method1}
\sigma(s)=\frac{C\beta} {s}(1+\frac{2mm_{\ast}}{s}) \frac{c_0}{(s-c_1)^4[\pi^2 + {\rm ln}^2(\frac{s}{\Lambda^2_{\rm QCD}})]^2},
\end{equation}
where the Coulomb factor $C$~\cite{BESIII:LcLc} parameterizes the electromagnetic interaction between the outgoing baryon and anti-baryon. This leads to a nonzero cross section near the threshold by canceling the velocity of the baryon in the c.m.~system, {\it i.e.}, $\beta$.
It takes $\beta=\sqrt{1-2(m^{2}+m_{\ast}^{2})/s+(m^{2}-m_{\ast}^{2})^2/s^{2}}$, where $m$ $(m_{\ast}$) denotes the nominal mass of the $\LamC$ ($\ALamCF$ or $\ALamCS$) baryon.
The free parameter $c_0$ is the normalization factor and $c_1$ indicates the contribution of potential resonant state.
The variable $\Lambda_{\rm QCD}$ is the QCD scale and is fixed to be 0.35 GeV.
After a few iterations, the Born cross sections converged, as tabulated in Table~\ref{tab::XS} and illustrated in Fig.~\ref{fig::Line_alpha} (a).

\linespread{1.1}
\begin{table*}[!htbp]
\centering
\caption{The %observed signal yields $N_{\rm sig}$, average detection efficiencies $\varepsilon =(\varepsilon_{ee}+\varepsilon_{\rm inte})/2$, ISR factors $f_{\rm ISR}$, and
Born cross sections of $\eeLLF+\rm{c.c.}$ and $\eeLLS+\rm{c.c.}$ at each energy point. The angular distribution parameter $\alpha_{\Lambda_{c}}$ and the form-factor ratios $\ffratio$ of $\eeLLS+\rm{c.c.}$.}
\label{tab::XS}
\renewcommand\arraystretch{1.2}
\setlength{\tabcolsep}{5.40mm}{
\begin{tabular}{c | c c | c c}
  \hline
  \hline
   Signal process & \multicolumn{2}{ c | }{$e^+e^-\to\LamC\bar{\Lambda}_{c}(2595)^{-}+\rm{c.c.}$}  & \multicolumn{2}{ c }{$e^+e^-\to\LamC\bar{\Lambda}_{c}(2625)^{-}$+\rm{c.c.}}\\ \hline
  $\sqrt{s}~(\mev)$  & 4918.0 & 4950.9 & 4918.0 & 4950.9 \\
  $N_{\rm sig}$   & $148\pm29$ & $216\pm27$ & $311\pm28$ & $552\pm47$ \\
 % $\varepsilon$ (\%) &$47.0\pm0.1$ & $46.8\pm0.1$ & $46.7\pm0.1$ & $46.8\pm0.1$ \\
 % $f_{\rm ISR}$ & 0.735 & 0.741 & 0.558 & 0.728 \\
  $\sigma$ (pb) & $15.6\pm3.1\pm0.9$ & $29.4\pm3.7\pm2.4$ & $43.4\pm4.0\pm4.1$ & $76.8\pm6.5\pm4.2$ \\
  $\alpha_{\Lambda_{c}}$ & - & - & $0.82\pm0.56\pm0.02$ & $-0.60\pm0.20\pm0.01$ \\
  $\ffratio$ & - & - &$5.95\pm4.07\pm0.15$&$0.94\pm0.32\pm0.02$\\
  \hline
  \hline
\end{tabular}
}
\end{table*}

%\vspace{-0.4em}
%\vspace{-0.8em}

Due to the limited statistics of the $\eeLLF+\rm{c.c.}$ signal yields, the angular distribution study is only performed for the $\eeLLS+\rm{c.c.}$ process.
Assuming that one-photon exchange dominates this process, the corresponding differential Born cross section is parametrized by three EMFFs, $|G_E|$, $|G_M|$, and $|G_C|$ (referred to as electric, magnetic, and Coulombic form factors, respectively)~\cite{Korner:1976hv,Devenish:1975jd}. It is written as
\begin{equation}
\frac{d\sigma}{d\!\cos\!\theta} \propto (1+\cos^2\!\theta)(|G_E|^2 + 3|G_M|^2) + \tau\cdot|G_C|^2\sin^2\!\theta,
\label{eq:xs}
\end{equation}
where $\tau=s/m_{\ast}^{2}$ with $m_{\ast}$ is the mass of $\LamCS$ and $\theta$ is the polar angle of the produced charmed baryon.
Since both energy points are close to the kinematic threshold, the effects from the ISR returned $\eeLLS+\rm{c.c.}$ events are negligible and the polar angle of the baryon is defined in the c.m. frame.
After parameterizing the polar angle distributions of the outgoing $\ALamCS$ and $\LamCS$ baryons with the function $f(\cos\theta) \propto (1+\alpha_{\Lambda_{c}}\!\cos^2\!\theta)$, the shape parameter $\alpha_{\Lambda_{c}}$ connects the EMFFs via
\begin{equation}
\label{eq:ff}
\frac{|G_E|^2 + 3|G_M|^2}{|G_C|^2} =\tau\cdot\frac{1+\alpha_{\Lambda_{c}}}{1-\alpha_{\Lambda_{c}}}.
\end{equation}

To determine $\alpha_{\Lambda_{c}}$, the $\cos\theta$ distributions of the $\ALamCS$ and $\LamCS$ polar angles are sliced into six and eight bins at $\sqrt{s}=4918.0$ and $4950.9~\mev$, respectively.
The yields in each $\cos\theta$ bin is determined with the same fit strategy to the $\Mrec$ spectrum as in the Born cross section measurement.
A bin-by-bin detection efficiency matrix, which takes into account the migration effect among different $\cos\theta$ bins, is used to correct the signal yields to the ones corresponding to generation level.
After that, $\alpha_{\Lambda_{c}}$ is extracted by fitting the obtained polar angle distributions with the $f(\cos\theta)$ function, and the resultant fit curves are presented in Fig.~\ref{fig::Line_alpha}(b).
Table~\ref{tab::XS} lists the obtained $\alpha_{\Lambda_{c}}$ and EMFF ratios.

The systematic uncertainties in the Born cross section measurements mainly arise from the tracking and PID efficiencies, $f_\text{\rm ISR}$, $f_\text{\rm VP}$, $\mathcal{L}$, $\mathcal{B}$, the signal MC modeling, and background descriptions.
The uncertainties from the tracking and PID efficiencies of the charged particles are investigated using the control samples of $J/\psi\to p\bar{p}\pi^+\pi^-$~\cite{Yuan:2015wga} and $J/\psi\to K^{0}_{S}K^{\pm}\pi^{\mp}$~\cite{BESIII:kk}, and 2.4\% is assigned for both c.m. energies.
The uncertainty in $f_\text{\rm ISR}$ is investigated using the approach described in Ref.~\cite{BESIII:LcLc},
which contains four aspects: different calculation algorithms~\cite{Ping:2016pms};
alternative input Born cross sections line shapes for the {\sc kkmc} generator;
the uncertainties in c.m. energy~\cite{BESIII:Lumi} and energy spread~\cite{Liu:2018ccy}.
The total uncertainties in $f_\text{\rm ISR}$ at $\sqrt s=4918.0~(4950.9)$ MeV are 2.2\% (3.5\%) for $\eeLLF+\rm{c.c.}$ and 7.8\% (1.3\%) for $\eeLLS+\rm{c.c.}$.
The uncertainty of $f_\text{\rm VP}$ is 0.5\% at both c.m. energies~\cite{WGRC:2010bjp,HC:1985gq}.
The uncertainty from the integrated luminosity is 0.5\% at both c.m.~energies, by studying the large-angle Bhabha scattering events~\cite{BESIII:Lumi}.
The uncertainty of $\mathcal{B}$ quoted from the PDG is 4.7\%~\cite{PDG:2022}.
Since the decay $\LamC\to pK^{-}\pi^{+}$ is well understood, the effect due to its MC modeling on the signal efficiency is negligible.
To estimate the uncertainty due to the signal MC modeling, the alternative signal MC samples are generated, in which the decay branching fractions of
$\LamCF$ and $\LamCS$, and the input $\alpha_{\Lambda_c}$ values are varied separately according to their total uncertainties.
In this procedure, the uncertainties in the decay branching fractions of $\LamCF$ and $\LamCS$ are quoted from PDG, and the ratio
between branching fraction of decay modes $\LamC\pi^+\pi^-$ and $\LamC\pi^0\pi^0$ is varied between 1.5:1 and 5:1.
The uncertainty in $\alpha_{\Lambda_c}$ for $\eeLLS+\rm{c.c.}$ is measured by this analysis, while for $\eeLLF+\rm{c.c.}$,
$\alpha_{\Lambda_c}=0$ is used to produce the alternative signal MC sample.
After comparing the Born cross sections given by the alternative and nominal signal MC samples, the associated uncertainties at
$\sqrt s=4918.0~(4950.9)$ MeV are 1.6\% (5.2\%) for $\eeLLF+\rm{c.c.}$ and 0.4\% (0.6\%) for $\eeLLS+\rm{c.c.}$.
The uncertainties in the background descriptions are studied by varying the background components and shapes, where the components $\ee \to \Sigma_{c}\bar{\Sigma}_{c}$ are removed and shape parameters of argus functions are obtained by fitting the inclusive MC samples, which are 0.9\% (0.1\%) for $\eeLLF+\rm{c.c.}$ and 0.7\% (0.8\%) for $\eeLLS+\rm{c.c.}$ at $\sqrt s=4918.0~(4950.9)$ MeV.
The total uncertainties in the Born cross sections measurements at $\sqrt s=4918.0~(4950.9)$ MeV are obtained by summing the individual contributions in quadrature, which are 6.0\% (8.2\%) for $\eeLLF+\rm{c.c.}$ and 9.4\% (5.5\%) for $\eeLLS+\rm{c.c.}$ .

The systematic uncertainties in the measurement of $\alpha_{\Lambda_c}$ arise from the binning effect, the efficiency correction, the signal MC modeling, and the background descriptions.
The systematic uncertainties arising from the binning effect are evaluated by analyzing the signal MC samples under the five-bin and ten-bin schemes.
The maximum differences of the resultant $\alpha_{\Lambda_{c}}$ from the ones obtained with the nominal binning schemes, which are 2.5\% and 0.7\% at $\sqrt{s}=4918.0$ and $4950.9~\mev$, respectively, are assigned as individual systematic uncertainties.
The uncertainty associated with the tracking and PID efficiencies are studied with the control samples mentioned above.
The corrected efficiency matrices are used to re-evaluate $\alpha_{\Lambda_c}$ and the resultant differences are taken as the systematic uncertainties, which are 0.1\% and 0.7\% at $\sqrt{s}=4918.0$ and $4950.9~\mev$, respectively.
A similar approach, as used in addressing the systematic uncertainties of Born cross sections, is applied to estimate the systematic uncertainties of $\alpha_{\Lambda_c}$ due to the signal MC modeling (the background descriptions), which are 0.4\% (0.1\%) and 0.6\% (0.6\%) at $\sqrt{s}=4918.0$ and $4950.9~\mev$, respectively.
The total uncertainties of $\alpha_{\Lambda_c}$ are obtained by summing the individual contributions in quadrature, which are 2.6\% and 1.3\% at $\sqrt{s}=4918.0$ and $4950.9~\mev$, respectively.
Accordingly, the uncertainties of the form-factor ratios, {\it i.e.}, $\ffratio$, are determined via the uncertainty propagation implied in Eq.~(\ref{eq:ff}).

In summary, the Born cross sections of the $\eeLLF+\rm{c.c.}$ and $\eeLLS+\rm{c.c.}$ processes are measured for the first time at $\sqrt{s}=4918.0$ and $4950.9~\mev$, by using $\ee$ collision data collected with the BESIII detector.
Non-zero cross sections very close to the production threshold are observed.
The measured Born cross sections of $\eeLLS+\rm{c.c.}$ above its production threshold are about $2\sim3$ times greater than those of $\eeLLF+\rm{c.c.}$, providing the similar behavior as semileptonic decays of $\Lambda_{b}^0$~\cite{CDF:2008hqh}, but different behavior from that in the hadronic decays of $\Lambda_{b}^0$. The improved measurements on both of the decays in $\Lambda_{b}^0$ to $\LamCF$ and $\LamCS$ and the productions via the electron-positron annihilation are expected in future, which could help us further understand the dynamics in the formation
of the excited baryons $\LamCF$ and $\LamCS$ from which it is possible to gain hints on the nature for these states.
%the internal structure of this two excited charmed baryons.
In addition, the angular distribution of the outgoing $\ALamCS$ and $\LamCS$
in the c.m. system is determined with the $f(\cos\theta) \propto (1+\alpha_{\Lambda_{c}}\!\cos^{2}\!\theta)$ parameterization.
The sign of $\alpha_{\Lambda_{c}}$ flips between these two energy points near the production threshold of $\LamC \bar{\Lambda}_{c}(2625)^- +\rm{c.c.}$.
The form-factor ratio $\ffratio$ is derived based on the $\alpha_{\Lambda_{c}}$ values for the first time.
This work opens a new window to explore the internal structure of the excited charmed baryons.
In the future, their internal structure is expected to be comprehensively understood via a fine scan of this energy region at $\ee$ collider with higher luminosity~\cite{Ablikim:2019hff}.

The BESIII Collaboration thanks the staff of BEPCII and the IHEP computing center for their strong support. This work is supported in part by National Key R\&D Program of China under Contracts No. 2020YFA0406300 and N0. 2020YFA0406400; National Natural Science Foundation of China (NSFC) under Contracts No. 11635010, No.~11735014, No.~11835012, No.~11935015, No.~11935016, No.~11935018, No.~11961141012, No.~12025502, No. 12035009, No. 12035013, No. 12061131003, No. 12192260, No. 12192261, No. 12192262, No. 12192263, No. 12192264, No. 12192265, No. 12221005, No. 12225509, No. 12235017 and No. 12005311; China Postdoctoral Science Foundation under Contracts No. 2019M662152, and No. 2020T130636; the Fundamental Research Funds for the Central Universities, University of Science and Technology of China under Contract No. WK2030000053; the Chinese Academy of Sciences (CAS) Large-Scale Scientific Facility Program; the CAS Center for Excellence in Particle Physics (CCEPP); Joint Large-Scale Scientific Facility Funds of the NSFC and CAS under Contract No. U1832207; CAS Key Research Program of Frontier Sciences under Contracts No. QYZDJ-SSW-SLH003 and No. QYZDJ-SSW-SLH040; 100 Talents Program of CAS; The Institute of Nuclear and Particle Physics (INPAC) and Shanghai Key Laboratory for Particle Physics and Cosmology; European Union's Horizon 2020 research and innovation programme under Marie Sklodowska-Curie grant agreement under Contract No. 894790; German Research Foundation DFG under Contracts No. 455635585, Collaborative Research Center Grants No. CRC 1044, No. FOR5327 and No. GRK 2149; Istituto Nazionale di Fisica Nucleare, Italy; Ministry of Development of Turkey under Contract No. DPT2006K-120470; National Research Foundation of Korea under Contract No. NRF-2022R1A2C1092335; National Science and Technology fund of Mongolia; National Science Research and Innovation Fund (NSRF) via the Program Management Unit for Human Resources \& Institutional Development, Research and Innovation of Thailand under Contract No. B16F640076; Polish National Science Centre under Contract No. 2019/35/O/ST2/02907; The Swedish Research Council; and U. S. Department of Energy under Contract No. DE-FG02-05ER41374.

\end{document}